\documentclass[12pt]{article}
\usepackage{amsmath,epsfig,amssymb,bbm,cite}

\newcommand{\jtheta}[1]{\vartheta \begin{bmatrix} #1 \end{bmatrix}}

\newcommand{\Z}{{\bf Z}}
\newcommand{\1}{{\mathbbm{1}}}

\renewcommand{\Im}{{\rm Im}}

\begin{document}

\title{\vbox{
\baselineskip 14pt
\hfill \hbox{\normalsize WU-HEP-09-04} \\
\hfill \hbox{\normalsize KUNS-2222}\\
\hfill \hbox{\normalsize YITP-09-49} } \vskip 2cm
\bf Magnetic flux, Wilson line and orbifold \vskip 0.5cm
}
\author{Hiroyuki~Abe$^{1,}$\footnote{email:
 abe@waseda.jp}, \
Kang-Sin~Choi$^{2,}$\footnote{email:
  kschoi@gauge.scphys.kyoto-u.ac.jp}, \
Tatsuo~Kobayashi$^{2,}$\footnote{
email: kobayash@gauge.scphys.kyoto-u.ac.jp} \ \\ and \
Hiroshi~Ohki$^{2,3,}$\footnote{email: ohki@scphys.kyoto-u.ac.jp
}\\*[20pt]
$^1${\it \normalsize
Department of Physics, Waseda University, Tokyo 169-8555, Japan} \\
$^2${\it \normalsize
Department of Physics, Kyoto University,
Kyoto 606-8502, Japan} \\
$^3${\it \normalsize 
Yukawa Institute for Theoretical Physics, Kyoto University, 
Kyoto 606-8502, Japan}
}

\date{}

\maketitle
\thispagestyle{empty}

\begin{abstract}
We study torus/orbifold models with magnetic flux and Wilson line 
background.
The number of zero-modes and their profiles depend on those 
backgrounds.
That has interesting implications from the viewpoint of 
particle phenomenology.
\end{abstract}

\newpage

\setcounter{page}{1}

\section{Introduction}

Extra dimensional field theory and string theory 
with magnetic fluxes can lead to interesting models~\cite{Manton:1981es,
Witten:1984dg,Bachas:1995ik,BDL,Blumenhagen:2000wh,
Angelantonj:2000hi,CIM,Troost:1999xn,Alfaro:2006is}.
Chiral theory can be realized as 4D effective field theory, 
because of magnetic flux background.
The number of zero modes, that is, the generation number, 
is determined by the magnitude of magnetic flux.
Their zero-mode profiles are non-trivially quasi-localized.
Such a behavior of zero-mode wavefunctions can lead to 
suppressed couplings when zero-modes are quasi-localized 
far away each other.
That would be useful to realize e.g. suppressed Yukawa couplings 
for light quarks and leptons.
On the other hand, when their localized points are close 
to each other, their couplings would be of ${\cal O}(1)$ and 
that would be useful to explain e.g. the top Yukawa coupling.
Furthermore, those localizing points on the torus background 
have a certain symmetry and it would become an origin of 
non-Abelian discrete flavor symmetries~\cite{Abe:2009vi}.\footnote{
Similar non-Abelian discrete flavor symmetries have been realized 
in heterotic orbifold 
models~\cite{Kobayashi:2004ya,Kobayashi:2006wq,Ko:2007dz}.} 
Furthermore, certain moduli can be stabilized by 
introducing magnetic fluxes~\cite{Antoniadis:2004pp}.
Thus, magnetized brane models have several phenomenologically 
interesting aspects.

In addition, magnetized D-brane models are T-dual of 
intersecting D-brane models~\cite{BDL,Blumenhagen:2000wh,
Angelantonj:2000hi,Aldazabal:2000dg,
Blumenhagen:2000ea,Cvetic:2001tj}.
(See for a review \cite{Blumenhagen:2005mu} and references therein.)
Within the framework of intersecting D-brane models, 
many interesting models have been constructed so far.

Magnetic backgrounds associated with  
orbifolds~\cite{Abe:2008fi,Abe:2008sx} and Wilson lines can also derive 
several interesting aspects and some of them have been 
studied.\footnote{Other backgrounds with magnetic fluxes were also 
studied~\cite{Conlon:2008qi,Marchesano:2008rg,Camara:2009xy}.}
Effects of Wilson lines on the torus with magnetic fluxes 
are gauge symmetry breaking and shift of wavefunction profiles.
Orbifolding is another way to realize a chiral theory.
For the same magnetic flux, the numbers of chiral zero-modes 
between the torus compactification and orbifold compactification 
are different from each other and zero-modes profiles are 
different~\cite{Abe:2008fi,Abe:2008sx}.
Adjoint matter fields remain massless on the torus with 
magnetic fluxes, those are projected out on the orbifold~\footnote{Within 
the framework of intersecting D-brane
  models, analogous results have been obtained by considering 
D6-branes wrapping rigid 3-cycles~\cite{Blumenhagen:2005tn}.}.
These differences lead to 
phenomenologically interesting aspects~\cite{Abe:2008sx}.
However, effects due to Wilson lines have not studied 
on the orbifold with the magnetic flux background.
Our purpose in this paper is to study more about 
these backgrounds such as consistency conditions, zero-mode profiles and 
phenomenological aspects of 4D effective theory.

This paper is organized as follows.
In section 2, we study 4D effective theory derived from 
the torus compactification with 
magnetic flux and Wilson line background.
Most of them are already known results.
(See e.g. \cite{CIM}.)
However, we reconsider phenomenological implications 
of Wilson lines on magnetized torus models.
In section 3, we study the orbifold background with 
magnetic fluxes and Wilson lines.
We study zero-modes under such a background 
and their phenomenological aspects.
Section 4 is devoted to conclusion and discussion.

\section{Magnetized torus models with Wilson lines}

\subsection{$T^2$ models}

Here, let us study 6D field theory with magnetic fluxes 
and Wilson lines.
The two extra dimensions are compactified on $T^2$, 
whose area and complex structure are denoted by 
$A$ and $\tau$.
We use the coordinates $y_m$ ($m=4,5$) for $T^2$, while 
$x_\mu$ ($\mu=0,\cdots, 3$) denote four dimensional uncompactified 
space-time, $R^{3,1}$.
Furthermore, we often use the complex coordinate, $z=y_4+\tau y_5$.
The boundary conditions on $T^2$ are represented by 
$z \sim z+1$  and $z \sim z + \tau$.

First, let us study $U(1)$ theory.
We consider the fermion field $\lambda (x,z)$ 
with $U(1)$ charge, $q$, and 
it satisfies the Dirac equation,
\begin{eqnarray}
\Gamma^M D_M \lambda(x,z) = \Gamma^M ( \partial_M -i q A_M )\lambda(x,z) = 0,
\end{eqnarray}
with $M=0,\cdots,5$, where $\Gamma^M$ denote the 6D gamma matrices and 
$A_M$ denote  $U(1)$ gauge vectors.
The fermion field $\lambda$, the vector fields $A_M$ ($M=(\mu, m)$)
are decomposed as 
\begin{eqnarray}
\lambda(x,z) &=& \sum_n \chi_n(x) \otimes \psi_n(z), 
\nonumber \\
A_\mu(x,z) &=& \sum_n A_{n,\mu}(x) \otimes \phi_{n,\mu}(z) , 
 \\
A_m(x,z) &=& \sum_n \varphi_{n,m}(x) \otimes \phi_{n,m}(z),
\nonumber
\end{eqnarray}
with $m=4,5$, 
where $A_{n,\mu}(x)$ and $\varphi_{n,m}(x)$ correspond to 
4D vector fields and scalar fields, respectively. 
Here, the modes with $n=0$ correspond to zero-modes, while 
the others correspond to massive modes. 
Since we concentrate on zero-modes, we omit the subscript 
corresponding to $n=0$.

We assume the following form of magnetic flux on $T^2$,
\begin{eqnarray}
F  ={\pi i \over \Im \tau} m \ (dz \wedge d \bar z), 
\end{eqnarray}
where $m$ is an integer~\cite{toron}.
Such magnetic flux can be obtained from the vector potential,
\begin{eqnarray}
A(z) ={\pi m \over \Im \tau} \Im (\bar z \ dz).
\end{eqnarray}
This form of the vector potential satisfies the 
following relations,
\begin{eqnarray}
A(z+1) &=& A(z) +{\pi m \over \Im \tau} \Im (dz), \\
A(z+\tau ) &=& A(z) +{\pi m \over \Im \tau} \Im (\bar \tau \ dz).
\end{eqnarray}
Furthermore, these can be represented as the following 
gauge transformations,
\begin{eqnarray}
A(z+1) = A(z) + d \chi_1, \qquad 
A(z+\tau ) = A(z) + d \chi_2,
\end{eqnarray}
where 
\begin{eqnarray}\label{eq:chi}
\chi_1 = {\pi m \over \Im \tau} \Im (z), \qquad 
\chi_2 = {\pi m \over \Im \tau} \Im (\bar \tau \ z).
\end{eqnarray}
Then, the fermion field $\psi(z)$ with the charge $q$ must 
satisfy 
\begin{eqnarray}
\psi (z+1) = e^{iq\chi_1(z)} \psi(z), \qquad 
\psi (z+\tau) = e^{iq\chi_2(z)} \psi(z).
\end{eqnarray}
Here and hereafter, we use the $U(1)$ charge normalization that 
all charges of matter fields are equal to integers and 
the minimum charge is $|q|=1$.
The internal part $\psi$ of the fermion zero-modes is 
two-component spinor,
\begin{eqnarray}
\psi = \left(
\begin{array}{c}
\psi_+  \\  \psi_-
\end{array}
\right),
\end{eqnarray}
and we choose the gamma matrix for $T^2$,
\begin{eqnarray}
\tilde \Gamma^1 = \left(
\begin{array}{cc}
0 & 1 \\ 1 & 0 
\end{array}
\right), \qquad \tilde \Gamma^2=\left(
\begin{array}{cc}
0 & -i \\ i & 0 
\end{array}
\right).
\end{eqnarray}
Then, the Dirac equations for zero-modes become 
\begin{eqnarray}
& & \left( \bar \partial + \frac{\pi qm}{2\Im (\tau)} z \right) 
\psi_+(z,\bar z) =0, \\
& & \left( \partial - \frac{\pi  qm}{2 \Im (\tau)} \bar z \right) 
\psi_- (z,\bar z) =0 .
\end{eqnarray}

When $qm >0$, the component $\psi_+$ has $M=qm$ independent 
zero-modes and their wavefunctions are  written as \cite{CIM}
\begin{eqnarray}
\Theta^{j,M}(z)=N_M e^{i \pi Mz\Im (z)/ \Im (\tau)} \vartheta 
\left[
\begin{array}{c}
j/M \\ 0
\end{array}
\right]
\left( Mz, M \tau \right),
\end{eqnarray}
where 
$N_M$ is a normalization factor, 
$j$ denotes the flavor index, i.e. 
$j=0,\cdots, M-1$ and 
\begin{eqnarray}
\vartheta \left[ 
\begin{array}{c}
a \\ b
\end{array} \right]
\left( \nu, \mu \right) 
&=& 
\sum_n \exp\left[ \pi i
   (n+a)^2 \mu + 2 \pi i (n+a)(\nu +b)\right],
\end{eqnarray}
that is, the Jacobi theta-function.
Note that $\Theta^{0,M}(z)=\Theta^{M,M}(z)$.
They satisfy the orthonormal condition,
\begin{equation} \label{orthre2}
 \int d^2z\ \Theta^{i,M}(z) \left( \Theta^{j,M}(z)  \right)^* =
  \delta_{ij}.
\end{equation}
Furthermore, for $qm >0$, the other component $\psi_-$ 
has no zero-modes.
As a result, we can realize a chiral spectrum.

On the other hand, when $qm <0$, 
the component $\psi_-$ has 
$|qm|$ independent zero-modes, but the other component 
$\psi_+$ has no zero-modes.

One of important properties of zero-mode wavefunctions is that 
we can decompose a product of two zero-mode wavefunctions 
as follows~\cite{Mu,Antoniadis:2009bg},
\begin{equation}\begin{split}\label{wvprod}
\Theta^{i,M_1}(z) \Theta^{j,M_2}(z) = 
&  \frac {N_{M_1} N_{M_2}}{N_{M_1+M_2}} 
\sum_{m \in \Z_{M_1+M_2}}  \Theta^{i+j+M_1m,M_1+M_2}(z)   \\
& \times \jtheta{{M_2 i - M_1 j + M_1 M_2 m \over M_1 M_2(M_1 +M_2)}
 \\ 0}(0,\tau_d M_1 M_2(M_1+M_2)). \\
\end{split} \end{equation}

Now, let us introduce Wilson lines.
The Dirac equations of the zero-modes are modified 
by the Wilson line background, $C = C_1 + \tau C_2$ as 
\begin{eqnarray}\label{eq:zero-mode-WL}
& & \left( \bar \partial + \frac{\pi  q}{2 \Im (\tau)}  (mz +C) \right) 
\psi_+(z,\bar z) =0, \\
& & \left( \partial - \frac{\pi  q}{2 \Im (\tau)}  (m\bar z+\bar C) \right) 
\psi_- (z,\bar z) =0, 
\end{eqnarray}
where $C_1$ and $C_2$ are real constants.
That is, we can introduce the Wilson line background, 
$C = C_1 + \tau C_2$ by replacing $\chi_i$ in 
(\ref{eq:chi}) as~\cite{CIM}
\begin{eqnarray}\label{eq:chi-WL}
\chi_1 = {\pi  \over \Im \tau} \Im (mz +C), \qquad 
\chi_2 = {\pi  \over \Im \tau} \Im (\bar \tau (mz + C)).
\end{eqnarray}
Because of this Wilson line, the number of 
zero-modes does not change, but their wavefunctions 
are replaced as 
\begin{eqnarray}\label{eq:WL-shift}
\Theta^{j,M}(z) \rightarrow \Theta^{j,M}(z+C/m).
\end{eqnarray}

In general, Yukawa couplings are computed by the overlap integral of three 
zero-mode profiles, $\psi_i(z)$, $\psi_j(z)$ and $\psi_k(z)$, 
\begin{eqnarray}\label{eq:yukawa}
y_{ijk} = g \int d^2z \psi_i(z) \psi_j(z) \psi_k(z),
\end{eqnarray}
where $g$ denotes the corresponding coupling in the 
higher dimensional theory.
Concretely, when Wilson lines are vanishing, the overlap integral of 
$\Theta^{i,M_1}(z)\Theta^{j,M_2}(z) (\Theta^{k,M_3}(z))^*$ 
for $M_3=M_1+M_2$ is given~\cite{CIM,DLMP} \footnote{
See also \cite{Russo:2007tc}.} 
\begin{eqnarray}\label{eq:yukawa-2}
& & \int d^2z \Theta^{i,M_1}(z)\Theta^{j,M_2}(z) (\Theta^{k,M_3}(z))^*
\nonumber \\ 
& & = \frac{N_{M_1}N_{M_2}}{N_{M_3}}
 \sum_{m \in \Z_{M_3}}  \delta_{i+j+M_1m,k}  
  \times \jtheta{{M_2 i - M_1 j + M_1 M_2 m \over M_1 M_2 M_3}
 \\ 0}(0,\tau M_1 M_2 M_3),
\end{eqnarray}
where the gauge invariance requires the third wavefuction must be 
$(\Theta^{k,M_3}(z))^*$ with the magnetic flux $M_3=M_1+M_2$, 
but not $\Theta^{k,M_3}(z)$.
Here we have used the product rule (\ref{wvprod}) and 
the orthogonality (\ref{orthre2}).
When we introduce non-vanishing Wilson lines, 
the overlap integral  is obtained as 
\begin{eqnarray}\label{eq:yukawa-wl}
& & \int d^2z \Theta^{i,M_1}(z+C/M_1)\Theta^{j,M_2}(z+C'/M_2) 
(\Theta^{k,M_3}(z+C''/M_3 ))^*\\ & & = 
\frac{N_{M_1}N_{M_2}}{N_{M_3}}
\sum_{m \in \Z_{M_3}}  \delta_{i+j+M_1m,k}  
 \times \jtheta{{M_2 i - M_1 j + M_1 M_2 m \over M_1 M_2 M_3}
 \\ 0}(M_2C-M_1C'),\tau M_1 M_2 M_3), \nonumber
\end{eqnarray}
where the gauge invariance requires $C''=C+C'$. 
Furthermore, by repeating the above procedure 
we can compute higher order couplings~\cite{Abe:2009dr}.

In Eqs.~(\ref{eq:yukawa-2}) and (\ref{eq:yukawa-wl}), 
the number of the Kronecker delta is defined modulo $M_3$ and 
the Kronecker delta leads to the selection rule for allowed 
couplings as 
\begin{eqnarray}
i+j-k= M_3 \ell -M_1m,
\end{eqnarray}
where $\ell, m$ are integers.
When $\gcd(M_1,M_2,M_3)=g$, the above constraint becomes 
\begin{eqnarray}\label{eq:Zg}
i+j = k, \qquad ({\rm~~mod}~~g~~).
\end{eqnarray}
That implies that we can define $Z_g$ charge for zero-modes and 
the allowed couplings are controlled by such a $Z_g$ 
symmetry~\cite{Abe:2009dr,Abe:2009vi}.\footnote{See for the same
  selection rule in intersecting D-brane 
models~\cite{CIMYukawa,Higaki:2005ie}.}
This $Z_g$ transformation can be written as~\cite{Abe:2009vi} 
\begin{eqnarray}\label{eq:Z}
Z = \left(
\begin{array}{ccccc}
1 & & & & \\
  & \rho & & & \\
  & & \rho^2 & & \\
  & &   & \ddots & \\
  &  &  &    & \rho^{g-1} 
\end{array}
\right),
\end{eqnarray}
where $\rho = e^{2\pi i /g}$.
Furthermore, 4D effective theory has a cyclic permutation 
symmetry 
\begin{eqnarray}
\Theta^{i,M_1} \rightarrow \Theta^{i+m n_1,M_1}, \quad
\Theta^{j,M_2} \rightarrow \Theta^{j+m n_2,M_2}, \quad
\Theta^{k,M_3} \rightarrow \Theta^{k+m n_3,M_3}, 
\end{eqnarray} 
where  $n_i=M_i/g$ and $m$ is a universal integer, that is, 
another $Z_g$ symmetry.
This $Z_g$ transformation can be written as~\cite{Abe:2009vi} 
\begin{eqnarray}\label{eq:C}
C = \left(
\begin{array}{cccccc}
0 & 1& 0 & 0 & \cdots & 0 \\
0  & 0 &1 & 0 & \cdots & 0\\
  &    &  & &\ddots & \\
1  &  0  & 0 & 0  & \cdots   & 0 
\end{array}
\right).
\end{eqnarray}
These two $Z_g$ symmetries are non-commutable and 
lead to non-Abelian flavor symmetry, 
$(Z_g \times Z_g) \rtimes Z_g$~\cite{Abe:2009vi}.
Its diagonal elements are written as $Z^m(Z')^n$, where 
\begin{eqnarray}\label{eq:Z-prime}
Z' = \left(
\begin{array}{ccc}
\rho & &  \\
    & \ddots & \\
    &    & \rho 
\end{array}
\right).
\end{eqnarray}
These symmetries are also available for higher order couplings.
Furthermore, when we consider vanishing Wilson lines, 
the $Z_2$ twist symmetry is enhanced by the symmetry, 
\begin{equation}
 \Theta^{i,M} \rightarrow \Theta^{M-i,M},
\end{equation}
and such  $Z_2$ can be written as 
\begin{eqnarray}\label{eq:cal-P}
{\cal P} = \left(
\begin{array}{ccccc}
1 &  \cdots & \cdots & \cdots & 0 \\
0 & \cdots & \cdots & 0 & 1 \\
0 & \cdots &0 & 1 & 0 \\
\vdots & &\rotatebox{70}{$\ddots$} & & \vdots \\
0 & 1 & 0 & \cdots & 0 
\end{array}
\right).
\end{eqnarray}
Then, the permutation 
symmetry is enhanced from $Z_g$ to $D_g$, and 
the total symmetry becomes 
$(Z_g \times Z_g) \rtimes D_g$.
For example, when $g=3$, we can realize 
$(Z_3 \times Z_3) \rtimes Z_3 = \Delta(27)$ and 
$(Z_3 \times Z_3) \rtimes D_3 = \Delta(54)$.

It would be useful to consider $U(1)_a \times U(1)_b$ theory from 
the phenomenological viewpoint.
We consider the fermion field $\lambda (x,z)$ 
with $U(1)_a \times U(1)_b$ charges, $(q_a,q_b)$.
We assume the following form of $U(1)_a$ magnetic flux on $T^2$,
\begin{eqnarray}\label{eq:magne-a}
F^a_{z \bar z} ={ \pi i \over \Im \tau} m_a, 
\end{eqnarray}
where $m_a$ is integer, but there is no magnetic flux in 
$U(1)_b$.
On top of that, we introduce Wilson lines $C^a$ and $C^b$ for 
$U(1)_a$ and $U(1)_b$, respectively.
The zero-mode equations are written as 
\begin{eqnarray}\label{eq:zero-mode-WL-ab}
& & \left( \bar \partial + \frac{\pi  }{2\Im (\tau)} 
 (q_a(m_az +C^a)+q_bC^b) \right) 
\psi_+(z,\bar z) =0, \\
& & \left( \partial - \frac{\pi  }{2\Im (\tau)} 
 (q_a(m_a\bar z +\bar C^a)+q_b\bar C^b) \right) 
\psi_- (z,\bar z) =0.
\end{eqnarray}
Then, the number of zero-modes is obtained as $M=q_a m_a $ 
and their wavefunctions are written as 
\begin{eqnarray}\label{eq:WL-shift-2}
\Theta^{j,M}(z+C/m_a),
\end{eqnarray}
where $C= C^a + C^bq_b/q_a$.
Here we give a few comments.
All of modes with $q_a=0$ become massive and there do not appear 
zero-modes with $q_a=0$.
For $q_a \neq 0$, zero-modes with $q_b=0$ appear 
and the number of zero-modes is independent of $q_b$.
Obviously, when we introduce Wilson lines $C^a$ and/or  $C^b$ 
without magnetic flux $F^a$, zero-modes do not appear.
The shift of wavefunctions depends on $1/m_a$ and 
the charge $q_b$.
Note that although $F^b=0$, 
Wilson lines $C_b$ and charges $q_b$ for $U(1)_b$ 
are also important \footnote{Wilson lines $C_b$ and charges $q_b$ for
  $U(1)_b$  are in a sense more important than 
Wilson lines $C_a$ and charges $q_a$ for $U(1)_a$, because  
the shift of wavefunctions (\ref{eq:WL-shift-2}) 
depends on $q_b$.}.

The above aspects of magnetic fluxes and Wilson lines 
are phenomenologically interesting.
We consider 6D super Yang-Mills theory with non-Abelian gauge group
$G$.
We introduce a magnetic flux $F^a$ along a Cartan direction of $G$.
Then, the gauge group breaks to $G'\times U(1)_a$ without reducing the rank.
Furthermore, there appear the massless fermion fields $\lambda'$ , 
which correspond to the gaugino fields for the broken gauge group
part and have the fundamental representation 
of $G'$ and non-vanishing $U(1)_a$ charge.
Furthermore, we introduce Wilson line along a Cartan direction of 
$G'$.
Then, the gauge group is broken to $G'' \times U(1)_a \times U(1)_b$ 
without reducing the rank.
The gaugino fields corresponding to the broken gauge part in $G'$ 
do not remain as massless modes, but they gain masses due to 
the Wilson line $U(1)_b$.
However, the fermion fields $\lambda'$ remain still massless 
with the same degeneracy.

\begin{figure}[t] \begin{center}
\includegraphics[height=3cm]{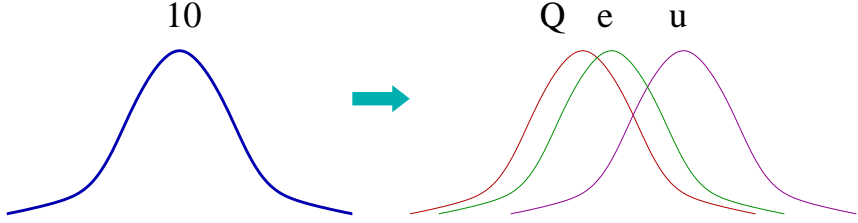}
\caption{Wavefunction splitting by Wilson lines}
\label{fig:WL-10}
\end{center} \end{figure}

Let us explain more on this aspect.
Suppose that we introduce magnetic fluxes in a model 
with a larger group $G$ such that they break 
$G$ to a GUT group like $SU(5)$ and 
this model include three families of matter fields 
like $10$ and $\bar 5$.
Their Yukawa couplings are computed by the overlap integral of three 
zero-mode profiles as Eq.~(\ref{eq:yukawa}).
We obtain the GUT relation among Yukawa coupling 
matrices when wavefunction profiles of matter fields in 
$10$ ($\bar 5$) are degenerate like Eq.~(\ref{eq:yukawa-2}).
Then, we introduce a Wilson line along $U(1)_Y$, which 
breaks $SU(5)$ to $SU(3) \times SU(2) \times U(1)_Y$.
Because of Wilson lines, $SU(5)$ gauge bosons except 
the $SU(3) \times SU(2) \times U(1)_Y$ gauge bosons become massive 
and the corresponding gaugino fields become massive.
However, three families of $10$ and $\bar 5$ matter fields remain massless.
Importantly, this Wilson line resolves the degeneracy of 
wavefunction profiles of left-handed quarks, right-handed up-sector 
quarks and right-handed charged leptons in $10$ and 
right-handed down-sector quarks and left-handed charged leptons 
in $\bar 5$ as Figure \ref{fig:WL-10}.
That  is, the GUT relation among Yukawa coupling matrices is 
deformed.
As an illustrating model, we study  the Pati-Salam model in 
the next subsection.

Here we study effects due to discrete values of Wilson lines 
such as $C=k\tau$ with $k=$ integer.
We find 
\begin{eqnarray}\label{eq:WV}
\Theta^{j,M}(z+k\tau/M) = e^{\pi i k \Im (\bar \tau z)/\Im(\tau)}
\Theta^{j+k,M}(z).
\end{eqnarray}
Thus, the effect of discrete Wilson lines $C=k\tau$ is 
to replace the $j$-th zero-mode by the $(j+k)$-th zero-mode
up to $e^{\pi i k \Im (\bar \tau z)/\Im(\tau)}$.
However, when we consider 3-point and higher order couplings, 
the gauge invariance requires that the sum of Wilson lines 
of matter fields should vanish, that is, $\sum_i k_i=0$ for 
allowed n-point couplings.
Thus, the part $e^{\pi i k \Im (\bar \tau z)/\Im(\tau)}$ is irrelevant to 
4D effective theory and 
the resultant 4D effective theory is the equivalent 
even when we introduce $C=k\tau$.
Similarly, introducing the Wilson lines $C=k$ with $k=$ integer leads 
to the equivalent 4D effective theory.

\subsection{Pati-Salam model}

As an illustrating model, we consider 
the Pati-Salam model.
We start with 10D N=1 $U(8)$ super Yang-Mills theory with the 
Lagrangian
\begin{eqnarray}
{\cal L} &=& 
-\frac{1}{4g^2}{\rm Tr}\left( F^{MN}F_{MN}  \right) 
+\frac{i}{2g^2}{\rm Tr}\left(  \bar \lambda \Gamma^M D_M \lambda
\right),
\end{eqnarray}
where $M,N=0,\cdots, 9$.
We compactify the extra 6 dimensions on 
$T^2_1\times T^2_2 \times T^2_3$, and we denote the complex coordinate 
for the $d$-th $T^2_d$ by $z^d$, where $d=1,2,3$.
Then, we introduce the following form of magnetic fluxes,
\begin{equation} \label{toronbg} \begin{split}
 F_{z^d \bar z^d} = {\pi i \over \Im \tau_d}  \begin{pmatrix}
   m_1^{(d)} \1_{4} & & \\
  & m_2^{(d)} \1_{2}  & \\ & & m_3^{(d)} \1_{2} 
\end{pmatrix}, \quad d=1,2,3,
\end{split} \end{equation}
in the gauge space, 
where $\1_{N}$ are the unit matrices of rank $N$, $m_i^{(d)}$ are
integers.
We assume that the above background preserves 4D N=1 supersymmetry (SUSY).
Here, we denote $M^{(d)}_{ij}=m^{(d)}_i - m^{(d)}_j$ and 
$M_{ij}=M^{(1)}_{ij}M^{(2)}_{ij}M^{(3)}_{ij}$.
This magnetic flux breaks the gauge group $U(8)$ to 
$U(4)\times U(2)_L \times U(2)_R $, that is the Pati-Salam gauge group 
up to $U(1)$ factors.
The gauge sector corresponds to 4D N=4 SUSY vector multiplet, that is, 
there are $U(4)\times U(2)_L \times U(2)_R $ N=1 vector multiplet and 
three adjoint chiral multiplets.
In addition, there appear bifundamental matter fields like  
$\lambda_{(4,2,1)}$, $\lambda_{(\bar 4,1,2)}$ and 
$\lambda_{(1,2,2)}$, and their numbers of zero-modes are equal to 
$M_{12}$, $M_{31}$ and $M_{23}$.
When $M_{ij}$ is negative, that implies their 
conjugate matter fields appear with the degeneracy $|M_{ij}|$.
The fields $\lambda_{(4,2,1)}$ and $\lambda_{(\bar 4,1,2)}$ 
correspond to left-handed and right-handed matter fields,
respectively, while $\lambda_{(1,2,2)}$ corresponds to 
up and down Higgs (higgsino) fields.
For example, we can realize three families by 
$M^{(d)}_{12}=(3,1,1)$ and $M^{(d)}_{31}=(3,1,1)$.
That leads to $|M_{23}|=0$ or 24.
At any rate, the flavor structure is determined by 
the first $T^2_1$ in such a model.
Explicitly, the zero-mode wavefunctions of both 
$\lambda_{(4,2,1)}$ and $\lambda_{(\bar 4,1,2)}$ 
are obtained as
\begin{eqnarray}
\Theta^{j,3}(z^1)\Theta^{1,1}(z^2)\Theta^{1,1}(z^3).
\end{eqnarray}
Their Yukawa matrices are constrained by the Pati-Salam 
gauge symmetry, that is, up-sector quarks, down-sector quarks,  
charged leptons and neutrinos have the same Yukawa matrices with 
Higgs fields.
Even with such a constraint, one could derive 
realistic quark/lepton masses and mixing angles, because 
this model has many Higgs fields and their vacuum expectation values 
generically break the up-down symmetry.

We introduce Wilson lines in $U(4)$ and $U(2)_R$ such that 
$U(4)$ breaks to $U(1)\times U(3)$ and $U(2)_R$ breaks 
$U(1) \times U(1)$.
Then, the gauge group becomes the standard gauge group 
up to $U(1)$ factors.
Furthermore, the profiles of left-handed quarks and 
leptons in $\lambda_{(4,2,1)}$ shift differently because of 
Wilson lines.
Similarly, right-handed up-sector quarks, down-sector quarks, 
charged leptons and neutrinos in $\lambda_{(\bar 4,1,2)}$ 
shift differently.
The flavor structure is determined by the first $T^2_1$.
Thus, when we introduce Wilson lines the second or third torus, 
the resultant Yukawa matrices are constrained by the 
$SU(4) \times SU(2)_L \times SU(2)_R$.
For example, we introduce Wilson lines on $T^2_2$.
Then, zero-mode profiles of 
quarks, $(Q,u,d)$ and leptons $(L,e,\nu)$ split as 
\begin{eqnarray}
Q&:& \Theta^{j,3}(z^1)\Theta^{1,1}(z^2+C^a)\Theta^{1,1}(z^3),
\nonumber \\
L&:& \Theta^{j,3}(z^1)\Theta^{1,1}(z^2-3C^a)\Theta^{1,1}(z^3),
\nonumber 
\\
u^c&:& \Theta^{j,3}(z^1)\Theta^{1,1}(z^2-C^a+C^b)\Theta^{1,1}(z^3), 
\nonumber \\
d^c&:& \Theta^{j,3}(z^1)\Theta^{1,1}(z^2-C^a-C^b)\Theta^{1,1}(z^3), \\
e^c&:& \Theta^{j,3}(z^1)\Theta^{1,1}(z^2+3C^a-C^b)\Theta^{1,1}(z^3), 
\nonumber \\
\nu^c&:& \Theta^{j,3}(z^1)\Theta^{1,1}(z^2+3C^a+C^b)\Theta^{1,1}(z^3),
\nonumber  
\end{eqnarray}
where $C^a$ and $C^b$ are the Wilson lines 
to break $U(4) \rightarrow U(3) \times U(1)$ and 
$U(2)_R \rightarrow U(1) \times U(1)$, respectively. 
Those Wilson lines just change the overall factors 
of Yukawa matrices, but ratios among elements in one Yukawa 
matrix 
do not change.
Also we can introduce Wilson lines along the same $U(1)$ directions 
as the magnetic fluxes (\ref{toronbg}), but they do not deform 
the up-down symmetry of Yukawa matrices, either.

On the other hand, when we introduce Wilson lines 
on the first $T^2_1$, the zero-mode wavefunctions
 split as 
\begin{eqnarray}
Q&:& \Theta^{j,3}(z^1+C^a/3)\Theta^{1,1}(z^2)\Theta^{1,1}(z^3),
\nonumber \\
L&:& \Theta^{j,3}(z^1-C^a)\Theta^{1,1}(z^2)\Theta^{1,1}(z^3),
\nonumber 
\\
u^c&:& \Theta^{j,3}(z^1-C^a/3+C^b/3)\Theta^{1,1}(z^2)\Theta^{1,1}(z^3), 
\nonumber \\
d^c&:& \Theta^{j,3}(z^1-C^a/3-C^b/3)\Theta^{1,1}(z^2)\Theta^{1,1}(z^3), \\
e^c&:& \Theta^{j,3}(z^1+C^a-C^b/3)\Theta^{1,1}(z^2)\Theta^{1,1}(z^3), 
\nonumber \\
\nu^c&:& \Theta^{j,3}(z^1+C^a+C^b/3)\Theta^{1,1}(z^2)\Theta^{1,1}(z^3). 
\nonumber  
\end{eqnarray}
In this case, the flavor structure is deviated from the 
$SU(4) \times SU(2)_L \times SU(2)_R$ relation, that is, 
mass ratios and mixing angles can change.
Also we can introduce Wilson lines $C^a$ to $T^2_2$ and $C^b$ to 
$T^2_1$.
Then we realize
\begin{eqnarray}
Q&:& \Theta^{j,3}(z^1)\Theta^{1,1}(z^2+C^a)\Theta^{1,1}(z^3),
\nonumber \\
L&:& \Theta^{j,3}(z^1)\Theta^{1,1}(z^2-3C^a)\Theta^{1,1}(z^3),
\nonumber 
\\
u^c&:& \Theta^{j,3}(z^1+C^b/3)\Theta^{1,1}(z^2-C^a)\Theta^{1,1}(z^3), 
\nonumber \\
d^c&:& \Theta^{j,3}(z^1-C^b/3)\Theta^{1,1}(z^2-C^a)\Theta^{1,1}(z^3), \\
e^c&:& \Theta^{j,3}(z^1-C^b/3)\Theta^{1,1}(z^2+3C^a)\Theta^{1,1}(z^3), 
\nonumber \\
\nu^c&:& \Theta^{j,3}(z^1+C^b/3)\Theta^{1,1}(z^2+3C^a)\Theta^{1,1}(z^3). 
\nonumber  
\end{eqnarray}

Indeed, this behavior is well-known in the intersecting 
D-brane models, which are T-duals of magnetized D-brane 
models.
In the intersecting D-brane side, introduction of 
Wilson lines corresponds to split D-branes.
By splitting D-branes, the gauge group breaks 
as $U(M+N) \rightarrow U(M)\times U(N)$, but 
the number of massless bi-fundamental modes does not change, 
although they decompose because of the gauge symmetry breaking.

\section{Orbifold models}

Here, we study orbifold models with magnetic fluxes and 
Wilson lines.
The $T^2/Z_2$ orbifold is constructed by identifying 
$z \sim -z$ on $T^2$.
We also embed the $Z_2$ twist into the gauge space as $P$.
%
%
Note that under the $Z_2$ twist, magnetic flux background is 
invariant.
That is, we have no constraint on magnetic fluxes due to 
orbifolding.  
Furthermore, zero-mode wavefunctions satisfy
\begin{eqnarray}
\Theta^{j,M}(-z) = \Theta^{M-j,M}(z).
\end{eqnarray}
Note that $\Theta^{0,M}(z) = \Theta^{M,M}(z)$.
Hence, the $Z_2$ eigenstates are written as \cite{Abe:2008fi}
\begin{eqnarray}\label{eq:wv-orbi}
\Theta^{j,M}_\pm(z) = \frac{1}{\sqrt 2} 
\left(\Theta^{j,M}(z) \pm \Theta^{M-j,M}(z) \right),
\end{eqnarray}
for $j\neq 0, M/2, M$.
The wavefunctions $\Theta^{j,M}(z)$ for $j=0, M/2$ are 
the $Z_2$ eigenstates with the $Z_2$ even parity.
Either of $\Theta^{j,M}_+(z)$ and $\Theta^{j,M}_-(z)$ 
is projected out by the orbifold projection.
Odd wavefunctions can also correspond to 
massless modes in the magnetic flux background, 
although on the orbifold without magnetic flux odd modes 
always correspond to massive modes, but not 
massless modes.
Before orbifolding, the number of zero-modes is equal to 
the magnetic flux $M$.
For example, we have to choose $M=3$ in order to realize 
the three families.
On the other hand, the number of zero-modes on the orbifold also 
depends on the boundary conditions under the $Z_2$ twist, 
even or odd functions.
For $M=$ even, 
the number of zero-modes with even (odd) functions 
are equal to $M/2+1$ $(M/2-1)$.
For $M=$ odd, 
the number of zero-modes with even and odd functions 
are equal to $(M+1)/2$ and $(M-1)/2$, respectively.
These results are shown in Table 1.
For example, when we choose even (odd) functions, the three families 
can be realized for $M=4$ and 5 (7 and 8).
Thus, we can obtain various three-family models in magnetized 
orbifold models and those have richer flavor structure than 
torus models with magnetic fluxes.
Yukawa couplings among 
$\Theta^{i,M_1}_{\pm}(z) \Theta^{j,M_2}_{\pm}(z) (\Theta^{k,M_3}_{\pm}(z))^*$ 
are computed by use of Eq.~(\ref{eq:yukawa-2}).

\begin{table}[t]
\begin{center}
\begin{tabular}{|c||c|c|}\hline
 & $M=$ even & $M=$ odd 
\\ \hline \hline 
even zero-modes & $M/2 +1$ &  $(M+1)/2 $ 
\\ \hline
odd zero-modes & $M/2 -1$  & $(M-1)/2 $
\\ \hline
\end{tabular}
\end{center}
\caption{The numbers of zero-modes for even and odd wavefunctions.}
\label{even-odd-zero-modes}
\end{table}




Now, let us introduce Wilson lines~\cite{Ibanez:1987xa}. 
We consider $U(1)_a \times SU(2)$ theory.
Then we introduce magnetic flux in $U(1)_a$ like Eq.~(\ref{eq:magne-a}).
In addition, we embed the $Z_2$ twist $P$ into the $SU(2)$ 
gauge space.
For example, we consider the $SU(2)$ doublet 
\begin{eqnarray}
\left(
\begin{array}{c}
\lambda_{1/2} \\ \lambda_{-1/2} 
\end{array}
\right),
\end{eqnarray}
with the $U(1)_a$ charge $q_a$.
We embed the $Z_2$ twist $P$ in the gauge space as 
\begin{eqnarray}\label{eq:P-su2}
P = \left( 
\begin{array}{cc}
0 & 1 \\
1 & 0 
\end{array}
\right),
\end{eqnarray}
for the doublet.
Obviously, we can diagonalize $P$ as 
$P'={\rm diag} (1,-1)$, 
if there is no Wilson line 
along the other $SU(2)$ directions.
However, we introduce a Wilson line along the 
Cartan direction of $SU(2)$, i.e, the following 
direction
\begin{eqnarray}
 \left( 
\begin{array}{cc}
1 & 0 \\
0 & -1 
\end{array}
\right),
\end{eqnarray}
in the $P$ basis.
Thus, we use the above basis for $P$.
For the $SU(2)$ gauge sector, there is no effect due to 
the magnetic flux.
Then, the situation is the same as one on 
the orbifold without magnetic flux.
The $SU(2)$ gauge group is broken completely, that is, 
all of $SU(2)$ vector multiplets become massive.

Before orbifolding, the $SU(2)$ is not broken and 
both $\lambda_{1/2}$ and $\lambda_{-1/2}$
have $M=q_am_a$ independent zero-modes, which we denote by 
$\Theta^{j,M}_{1/2}(z)$ and $\Theta^{j,M}_{-1/2}(z)$, respectively.
Here, we have put the indices, $1/2$ and $-1/2$ in order to 
make it clear that they correspond to $\lambda_{1/2}$ and
$\lambda_{-1/2}$, respectively.
However, the form of wavefunctions are the same, i.e. 
$\Theta^{j,M}_{1/2}(z) = \Theta^{j,M}_{-1/2}(z)$.
When we impose the orbifold boundary conditions with 
the above $P$ in (\ref{eq:P-su2}), 
the zero-modes on the orbifold without Wilson lines are written as 
\begin{eqnarray}\label{eq:Z2-state}
\frac{1}{\sqrt 2} \left(\Theta^{j,M}_{1/2}(z) 
+ \Theta^{M-j,M}_{-1/2}(z) \right),
\end{eqnarray}
for $j=0,\cdots, M-1$.
Note that there are $M$ independent zero-modes.
It may be useful to explain remaining zero-modes in the basis 
for $P'$.
Before orbifolding, both $\lambda'_{1/2}$ and $\lambda'_{-1/2}$
have $M=q_am_a$ independent zero-modes in the basis for $P'$.
Then by orbifolding with $P'$, even modes corresponding 
to $\Theta^{j,M}_+ (z)$ remain for $\lambda'_{1/2}$, 
while  $\lambda'_{-1/2}$ has only odd modes $\Theta^{j,M}_- (z)$.
Their total number is equal to $M$.

Then, we introduce the Wilson lines along the 
Cartan direction in the basis for $P$.
The corresponding zero-mode wavefunctions are shifted as 
\begin{eqnarray}\label{eq:wf-cWL}
\frac{1}{\sqrt 2} \left(\Theta^{j,M}_{1/2}(z+C^b/2M) 
+ \Theta^{M-j,M}_{-1/2}(z-C^b/2M) \right),
\end{eqnarray}
for $j=0,\cdots, M-1$, where $C^b$ is a continuous parameter.
Note that $\lambda_{1/2}$ and $\lambda_{-1/2}$
have opposite charges under the $SU(2)$ Cartan.
Then, their wavefunctions are shifted to opposite directions   
by the same Wilson lines $C^b$
as $\Theta^{j,M}_{1/2}(z+C^b/2M) $ and $\Theta^{j,M}_{-1/2}(z-C^b/2M)$.
We can also consider the $Z_2$ twist $P$ in the doublet such that 
the following wavefunction 
\begin{eqnarray}\label{eq:wf-cWL-}
\frac{1}{\sqrt 2} \left(\Theta^{j,M}_{1/2}(z+C^b/2M) -
 \Theta^{M-j,M}_{-1/2}(z-C^b/2M) \right),
\end{eqnarray}
remains.

For comparison, we study another dimensional representations, 
e.g. a triplet
\begin{eqnarray}
\left(
\begin{array}{c}
\lambda_{1} \\ \lambda_{0} \\\lambda_{-1} 
\end{array}
\right),
\end{eqnarray}
with the $U(1)_a$ charge $q_a$.
Suppose that we embed the $Z_2$ twist $P$ in the three dimensional 
gauge space as 
\begin{eqnarray}\label{eq:P-su2-3}
P = \left( 
\begin{array}{ccc}
0 & 0 & 1 \\
0 & 1 & 0 \\
1 & 0 & 0
\end{array}
\right),
\end{eqnarray}
for the triplet.
Then, zero-modes on the orbifold are written as 
\begin{eqnarray}
 & &  \Theta^{j,M}_{1}(z) 
+ \Theta^{M-j,M}_{-1}(z) , \nonumber \\
 & &\Theta^{j,M}_{0}(z) 
+ \Theta^{M-j,M}_{0}(z) ,
\end{eqnarray}
up to the normalization factor $1/\sqrt{2}$.
The former corresponds to $\lambda_{1}$ and $\lambda_{-1}$ and 
there are $M$ zero-modes.
The latter corresponds to $\lambda_0$ and there are 
$(M/2+1)$ zero-modes and $(M+1)/2$ zero-modes when $M$ is even and odd, 
respectively.
When we introduce the continuous Wilson lines along the Cartan 
direction, these zero-modes shift as 
\begin{eqnarray}
 & & \Theta^{j,M}_{1}(z+C_b/M) 
+ \Theta^{M-j,M}_{-1}(z-C_b/M) , \nonumber \\
 & &\Theta^{j,M}_{0}(z) 
+ \Theta^{M-j,M}_{0}(z) .
\end{eqnarray}

We can extend the above analysis to larger gauge groups.
Here, we show a rather simple example.
We consider $U(1)_a \times SU(3)$ theory with 
the magnetic flux in $U(1)_a$ like Eq.~(\ref{eq:magne-a}).
Then, we consider the $SU(3)$ triplet, 
\begin{eqnarray}
\left(
\begin{array}{c}
\lambda_{0} \\ \lambda_{1/2} \\ \lambda_{-1/2} 
\end{array}
\right),
\end{eqnarray}
with the $U(1)_a$ charge $q_a$, where 
the subscripts $(0,1/2,-1/2)$ denote the $U(1)_b$ charge 
along one of $SU(3)$ Cartan directions.
Now, we embed the $Z_2$ twist $P$ in the gauge space as 
\begin{eqnarray}\label{eq:P-su3}
P = \left( 
\begin{array}{ccc}
1 & 0 & 0 \\
0 & 0 & 1 \\
0 & 1 & 0
\end{array}
\right),
\end{eqnarray}
for the triplet.
In addition, we introduce the Wilson line $C^b$ along 
the $U(1)_b$ direction.
The gauge group is broken as $SU(3) \rightarrow U(1)$.\footnote{
This remaining $U(1)$ symmetry might be anomalous.
If so, the remaining $U(1)$ would also be broken by 
the Green-Schwarz mechanism.}
There are $M$ zero-modes for linear combinations of 
$\lambda_{1/2}$ and $\lambda_{-1/2}$ with the wavefunctions,
\begin{eqnarray} 
\Theta^{j,M}_{1/2}(z+C^b/2M) +
 \Theta^{M-j,M}_{-1/2}(z-C^b/2M),
\end{eqnarray}
 up to the normalization factor.
Also, the zero-modes for $\lambda_{0}$ are written as 
\begin{eqnarray} 
\Theta^{j,M}_{0}(z) +
 \Theta^{M-j,M}_{0}(z),
\end{eqnarray}
up to the normalization factor.
The number of zero-modes is equal to 
$(M/2+1)$ and $(M+1)/2$ when $M$ is even and odd, respectively.
Thus, the situation is almost the same as the 
above $SU(2)$ case with the triplet.
Although the above example is rather simple, 
we can consider various types of breaking for larger groups.
For example, when the gauge group includes 
two or more $SU(2)$ subgroups, 
we could embed the $Z_2$ twist in two of $SU(2)$'s and 
introduce independent Wilson lines along their Cartan directions.
Similarly, we can investigate such models and other types of  
various embedding of $P$ and Wilson lines.

In section 2.2, we have considered 10D theory on $T^6$.
Also, we can consider the $T^6/Z_2$ orbifold, where the $Z_2$ 
twist acts e.g. 
\begin{eqnarray}
Z_2 \ : \ z_1 \rightarrow -z_1, \qquad 
z_2 \rightarrow -z_2, \qquad 
z_3 \rightarrow z_3.
\end{eqnarray}
For $T^2_1$ and $T^2_2$, we can introduce the type 
of Wilson lines, which we have considered in this section, 
while for $T^2_3$ we can introduce the type of Wilson lines, 
which are considered in the previous section.
Then, we have a richer structure of models on 
the $T^6/Z_2$ orbifold.
Furthermore, we could consider another independent $Z'_2$ twist 
as 
\begin{eqnarray}
Z_2 \ : \ z_1 \rightarrow -z_1, \qquad 
z_2 \rightarrow z_2, \qquad 
z_3 \rightarrow -z_3,
\end{eqnarray}
on the $T^6/(Z_2 \times Z'_2)$ orbifold.
In this case, we can consider another independent 
embedding $P'$ of $Z'_2$ twist on the gauge space.
Using these two $Z_2$ twist embedding and Wilson lines, we 
could construct various types of models.
For example, when the gauge group includes two or 
more $SU(2)$ subgroups, we could embed $P$ on one of $SU(2)$ 
and $P'$ on other $SU(2)$ and introduce independent 
Wilson lines along their Cartan directions.
Other various types of model building would be possible.
Thus, it would be interesting to study such model building 
elsewhere.

Finally, we comment on the flavor symmetry.
Yukawa couplings as well as higher order couplings can be  
computed by use of Eq.~(\ref{eq:yukawa-wl}).
The orbifolding without Wilson lines is a procedure to 
choose eigenstates for ${\cal P}$ (\ref{eq:cal-P}).
Thus, there remains the flavor symmetry, which commutes with 
${\cal P}$.
The $Z_g'$ symmetry (\ref{eq:Z-prime}) is commutable.
The $Z_g$ symmetry (\ref{eq:Z}) is not commutable for $g=$ odd.
However, when $g=$ even, the $Z_2$ symmetry, which is generated by 
$Z^{g/2}$ is commutable with ${\cal P}$ and another $Z_2$ symmetry, 
which is generated by $C^{g/2}$, is also commutable with ${\cal P}$.
For example, when $g=4$, the generators, $Z'$, $Z^2$ and $C^2$ are 
written as 
\begin{eqnarray}
Z' = \left(
\begin{array}{ccc}
i & &  \\
    & \ddots & \\
    &    & i 
\end{array}
\right), \qquad 
Z^2 = \left(
\begin{array}{cccc}
1 & & &  \\
  & -1 & &  \\
  & & 1 &  \\
  &  &      & -1 
\end{array}
\right), \qquad 
C^2 = \left(
\begin{array}{cccc}
0 & 0& 1 & 0  \\
0  & 0 &0 & 1 \\
1  & 0   & 0 &0  \\
0  &  1  & 0 &  0 
\end{array}
\right).
\end{eqnarray}
Here, the $Z^2$ and $C^2$ generators also commute with each other.
Similarly, when $g/2 =$ even,  
the unbroken flavor symmetry would be obtained as
$Z_g \times Z_2 \times Z_2\times Z_2$.\footnote{
It is interesting to break non-Abelian flavor symmetries 
to Abelian symmetries by orbifolding~\cite{Kobayashi:2008ih}.}
On the other hand, when $g/2=3$,   
the generators, $Z'$, $Z^3$ and $C^3$ are 
written as 
\begin{eqnarray}
& & Z' = \left(
\begin{array}{ccc}
\rho & &  \\
    & \ddots & \\
    &    & \rho 
\end{array}
\right), \nonumber \\
& & Z^3 = \left(
\begin{array}{cccccc}
1 & & & & &  \\
  & -1 & & & &  \\
  & & 1 &  & & \\
 & &   & -1 & &  \\
  & & & & 1 &  \\
  &  & & &       & -1 
\end{array}
\right), \qquad 
C^3 = \left(
\begin{array}{cccccc}
0 & 0 & 0& 1 & 0 & 0 \\
0  & 0 & 0 &0 & 1 & 0 \\
0  & 0 & 0 &0 & 0 & 1 \\
1  & 0   & 0 &0 & 0 & 0 \\
0  &  1  & 0 &  0 & 0 & 0 \\
0  &  0  & 1 &  0 & 0 & 0 \\
\end{array}
\right),
\end{eqnarray}
where $\rho = e^{\pi i/3}$.
Here, the $Z^3$ and $C^3$ generators do not 
commute with each other.
Thus, unbroken flavor symmetries are non-Abelian.
Similarly, when $g/2=$ odd, non-Abelian discrete flavor 
symmetries would remain.

When we introduce the above Wilson lines, 
the $SU(2)$ gauge symmetry is broken at the same time as  
the $Z_g$ symmetry breaking for (\ref{eq:Z}).
Thus, we may expect that some non-trivial linear combinations 
of broken $Z_g$ and $SU(2)$ would remain.
However, only the $Z_2$ symmetry, which is already included above, 
seems to remain e.g. in the states (\ref{eq:Z2-state}).
When we consider more complicated models, 
a new type of flavor symmetries, which are linear 
combinations of broken flavor symmetries and gauge symmetries, 
may remain.
Hence it would be interesting to investigate such models.

\section{Conclusion and discussion}

We have studied torus/orbifold models with magnetic fluxes and Wilson lines.
These backgrounds lead to various different aspects like 
the number of zero-modes, their profiles, breaking patterns of 
flavor symmetries, etc.
Using these backgrounds, it would be quite interesting to 
construct concrete models.
We would study them elsewhere.

In addition to continuous Wilson lines, 
we can introduce discrete Wilson lines 
on the orbifold without magnetic fluxes, 
which break the gauge group without reducing its rank.
It is quite important to study the possibility for 
introducing discrete Wilson lines 
in the magnetic background and 
study their phenomenological implications.
Furthermore, it is also important to analyze 
(systematically) which types of backgrounds 
and boundary conditions are possible in generic case.

\subsection*{Acknowledgement}

The authors would like to thank R.~Maruyama, M.~Murata, Y.~Nakai
and M.~Sakai for useful discussions.
H.~A. is supported in part by the Waseda University Grant for 
Special Research projects No.~2009A-854.
K.-S.~C., T.~K. and H.~O. are supported in part by the Grant-in-Aid for 
Scientific Research No.~20$\cdot$08326, No.~20540266 and
No.~21$\cdot$897 from the 
Ministry of Education, Culture, Sports, Science and Technology of Japan.
T.~K. is also supported in part by the Grant-in-Aid for the Global COE 
Program "The Next Generation of Physics, Spun from Universality and 
Emergence" from the Ministry of Education, Culture,Sports, Science and 
Technology of Japan.

\end{document}